# Distinct spatial characteristics of industrial and public research collaborations: Evidence from the 5[th] EU Framework Programme

Thomas Scherngell[*] and Michael Barber

[*]Corresponding author,
Department of Technology Policy, Austrian Research Centers GmbH – ARC,
Donau-City-Strasse 1, A-1220 Vienna, Austria
Email: thomas.scherngell@arcs.ac.at

**Abstract.** This study compares the spatial characteristics of industrial R&D networks to those of public research R&D networks (i.e. universities and research organisations). The objective is to measure the impact of geographical separation effects on the constitution of cross-region R&D collaborations for both types of collaboration. We use data on joint research projects funded by the 5[th] European Framework Programme (FP) to proxy cross-region collaborative activities. The study area is composed of 255 NUTS-2 regions that cover the EU-25 member states (excluding Malta and Cyprus) as well as Norway and Switzerland. We adopt spatial interaction models to analyse how the variation of cross-region industry and public research networks is affected by geography. The results of the spatial analysis provide evidence that geographical factors significantly affect patterns of industrial R&D collaboration, while in the public research sector effects of geography are much smaller. However, the results show that technological distance is the most important factor for both industry and public research cooperative activities.

# 1 Introduction

Today it is widely agreed that innovation, knowledge creation and the diffusion of new knowledge are the key vehicles for sustained economic growth of firms, industries or regions, and, thus, are essential for achieving sustained competitive advantage in the economy. In theoretical models of New Growth Theory, it is argued that geographical space is crucial for innovation since important parts of new knowledge have some degree of tacitness that is embedded in the routines of individuals and organisations (see, for example, Romer 1990). It is assumed that sustained long-run growth emanates from geographically localized knowledge spillovers and knowledge diffusion. These theoretical considerations have been followed by several empirical studies that have investigated the spatial dimension of knowledge flows, but from its beginning the measurement of such phenomena has faced numerous problems. "Knowledge flows are invisible, they leave no paper trail by which they may be measured and tracked," states Krugman (1991, pp. 153), pointing to difficulties in finding data on knowledge diffusion with respect to the geographical and temporal location of such processes. However, the pioneering study of Jaffe et al. (1993) demonstrates that knowledge flows do sometimes leave a paper trail in the form of patent citations. Their study produces evidence for the localisation hypothesis of knowledge diffusion processes.

The following years of empirical research on the geography of knowledge flows have been characterized by the development of new indicators and the integration of new econometric and statistical techniques (see, for instance, Maurseth and Verspagen 2002, Fischer, Scherngell and Jansenberger 2006). One major recent novelty of this research stream is the possibility to use formalised R&D collaborations as a proxy for knowledge flows, since during the recent past such networks have become increasingly formalised and have, thus, received greater institutional visibility. The usage of such indicators can be justified by the increasing importance of collaborative activities and networks in the process of knowledge creation over the past years. The increasing complexity of innovation processes makes it inevitable for organisations to tap external sources of knowledge. These changes set up new questions concerning the role of geographical space for innovation and knowledge diffusion, since the arrangement of R&D collaboration networks may modify the spatial diffusion of knowledge.

From this perspective, Autant-Bernard et al. (2007a) suggest that the geographical dimension of knowledge diffusion warrants further attention by analysing phenomena such as R&D collaborations. The study of Maggioni et al. (2007) is one important recent contribution in this context by focusing on knowledge-based relational phenomena directly linked to collaborative activities in the form of co-patent applications and joint participation in European research networks. Their study provides evidence that a region's knowledge production is not only positively influenced by knowledge production activities of regions that are located nearby in geographical space, but also by regions that are relationally close as captured by joint participation in European research networks. Scherngell and Barber (2009) investigated the geography of R&D collaborations across European regions by using spatial interaction modelling techniques. They used data on joint FP5 research projects as a proxy for cross-region collaboration activities. Their study provides evidence that geographical distance significantly affects patterns of cross-region R&D collaborations in Europe[1]. However, they show that technological proximity matters more than geographical proximity. In this paper, we report on a study that is similar in spirit, but expands on the earlier approach by distinguishing between intra-industry and intra-public-research (i.e. universities and research organisations) R&D collaborations. It is our working assumption that the results obtained in Scherngell and Barber (2009) may differ across these types of organisation. Supporting this assumption, Almendral et al. (2008) have shown that the network properties of universities and industries in FP5 differ markedly from one another.

The objectives of this study are, *first*, to identify collaboration patterns of cross-region intra-industry and intra-public-research R&D collaborations in Europe, and, *second*, to measure separation effects on the constitution of cross-region intra-industry and intra-public-research R&D collaborations in Europe. In particular, we are interested to explore whether different kinds of geographical space – such as physical distance between regions or existence of country borders between regions – show different effects on these types of collaboration. R&D collaborations in the industry and the public research sector are captured by data on research projects of the $5^{th}$ EU Framework Programme (FP5). We take as a working hypothesis that FP5 research projects serve as a proxy for cross-region collaborative R&D activities in Europe and shift attention to regions as units of analysis by using aggregated R&D collaboration data at the regional level. As in Scherngell and Barber (2009), we adopt a Poisson spatial interaction modelling perspective to estimate the separation effects for cross-region industry and public research networks. The study area is composed of 255 NUTS-2 regions that cover the 25 pre-2007 EU member-states (Malta and Cyprus excluded), as well as Norway and Switzerland.

---

[1] Other recent contributions that investigate geographical aspects of R&D collaborations involve the studies of Constantelou et al. (2004), Autant-Bernard et al. (2007b), Maggioni et al. (2007) Maggioni and Uberti (2007) and Hoekman et al. (2008).

The current study departs from previous literature in at least three major respects. *First*, the distinction in the empirical investigation between industry and public research networks established within the European FPs is an original contribution of the current study. *Second*, we use regions as units of analysis in order to investigate patterns of EU FP collaborations, in contrast to most other studies that use organisations as observational units (see, for instance, Almendral et al. 2008, Roediger-Schluga and Barber 2008). This is an appropriate choice since we shift emphasis on the spatial characteristics of industry versus public research FP collaborations. *Third*, by using EU FP data aggregated to 255 regions of the 25 pre-2007 EU member-states including Norway and Switzerland, our study extends the geographic coverage in comparison to many previous studies (see, for instance, Maggioni et al. 2007).

The remainder of this paper is organised as follows. *Section 2* presents the theoretical background and outlines the main hypotheses for the empirical study. *Section 3* describes the European Framework Programmes in more detail, before *Section 4* introduces the empirical setting and the data used in the current study, accompanied by some descriptive statistics. *Section 5* explores the spatial structure of industry and public research R&D networks across Europe by using network analysis techniques in combination with GI tools. *Section 6* describes the spatial interaction modelling perspective, followed by the estimation results presented in *Section 7*. *Section 8* concludes with a summary of the main results, some policy implications and a short outlook.

## 2  Theoretical Background and Main Hypotheses

Today, it is widely recognized that interactions between firms, universities and research organisations are a *sine qua non* condition for successful innovation in the current era of the knowledge-based economy (see, for example, OECD 1992). Many recent empirical studies highlight that knowledge production involves an increasing number of actors who are more interconnected to each other (see Autant-Bernard et al. 2007b). Pavitt (2005) notes that the growing complexity of technology and the existence of converging technologies are key reasons for this development. In particular, firms have expanded their knowledge bases into a wider range of technologies (Granstrand 1998), increasing the need for distinct types of knowledge, so firms must learn how to integrate new knowledge into existing products or production processes (Cowan 2004). It may be difficult to develop this knowledge alone or acquire it via the market. Thus, firms form cooperative arrangements with other firms, universities or research organisations that already have this knowledge, to access it more rapidly. The fundamental importance of interactions and networks for innovations is also reflected in the various systems of innovation concepts (see Lundvall 1992, among others). In this conception, sources of innovation are often established between firms, universities, suppliers and customers. Network arrangements create incentives for interactive organisational learning, leading to faster knowledge diffusion within the innovation system and stimulating the creation of new knowledge or new combinations of existing knowledge. In particular, network arrangements are useful in the presence of uncertainty and complexity, such as in innovation processes. Participation in innovation networks reduces the degree of uncertainty and provides fast access to different kinds of knowledge, in particular tacit knowledge (see, for example, Kogut 1988).

The analysis in the current paper focuses on measuring separation effects on the constitution of cross-region intra-industry and cross-region intra-public-research R&D collaborations in Europe. We distinguish three streams in the theoretical and empirical literature, which suggests focusing on three main hypotheses[2]:

---

[2] In this context, it is worth emphasising that this paper lies in the research tradition that investigates the geographical dimension of knowledge flows. Another possibility would be to take a social network or matching games perspective (see, for instance, Goyal 2007, Tabellini 2008).

*First*, New Economic Geography and New Trade Theory stress the importance of geographic proximity and spatial location for knowledge flows and knowledge production activities. Krugman (1991) argues that knowledge flows among knowledge producing agents may be geographically bounded since important parts of new knowledge have some degree of tacitness. Though the cost of transmitting codified knowledge may be invariant to distance, presumably the cost of transmitting non-codified knowledge across geographic space rises with geographic distance (see Audretsch and Feldman 1996). There is a large body of empirical literature that lies in the research tradition investigating the geographic localisation of knowledge flows[3]. From this background we hypothesise that the variation of cross-region R&D collaboration intensities significantly depends on the spatial distance between any two regions. To address this question we estimate the effect of geographical distance between regions on their collaboration probability as well as the existence of spatial clustering effects.

*Second*, theoretical literature on social interactions, social networks and collaboration behaviour stress the importance of the time and money required to engage in collaborations (see, for instance, Boschma 2005, Ponds et al. 2007, Hoekman et al. 2008). Gertler (1995) and Edquist and Johnson (1997) point to the importance of institutional and cultural proximity for collaboration. It is argued that costs of research collaboration increase with institutional and cultural distance, while a common institutional environment decreases costs. Thus, we hypothesise that the probability of collaboration between any two regions decreases if they are located in different institutional/cultural environments, such as different countries or different language areas. In this context it is worth emphasising that the FPs are intended to overcome institutional and cultural barriers between the EU member states (see CORDIS 2006).

---

[3] An important recent contribution is the study of Maggioni et al. (2007) that confirms the significance of spatial effects for the constitution of collaborative knowledge production activities. Other important studies include Jaffe, Trajtenberg and Henderson (1993), Katz (1994), Anselin et al. (1997), Breschi and Lissoni (2001), Bottazi and Peri (2003), Maurseth and Verspagen (2004), Fischer et al. (2006), LeSage et al. (2007).

*Third*, another important issue in relevant empirical and theoretical literature is the extent to which inter-industry or inter-technology knowledge flows take place (see Lucas 1988). If knowledge flows mainly within sectors, the impact of industrial specialisation may be different than if knowledge flows easily between economic and technological fields. According to previous empirical studies that investigate the impact of economic and technological distance on knowledge flows (see, for instance, Maurseth and Verspagen 2002), we hypothesise that the probability of cross-region collaboration activities increases by economic similarity and technological proximity.

## 3  The European Framework Programmes

In Europe, the prime examples of instruments that foster collaborative R&D activities are the Framework Programmes (FPs) on Research and Technological Development. By means of this initiative, the EU has co-funded thousands of trans-national, collaborative R&D projects. The main objectives of the instrument from the background of a European technology policy view are to integrate national and regional research communities and to coordinate national policies. Empirical studies such as the one of Breschi and Cusmano (2004) point to the establishment of a pan-European network of firms, universities, public research organisations, consultants and government institutions performing joint research in different thematic fields.

Implementation of the EU FPs began in 1984; the current seventh programme has begun in 2007 and will run until 2013[4]. Since the launch of the FPs in 1984, EU institutions have focused funding on multidisciplinary research at a trans-national level. Over the years, different thematic aspects and issues of the European scientific landscape have been addressed by the FPs, though the main emphasis has shifted more and more towards the establishment of an integrated European Research Area. In this sense, the European technology policy aims to promote technological competitiveness, while at the same time it is meant to ensure cohesion. Projects to be funded by the FPs must fulfil a number of requirements that shape the particular outset of each FP. One of the key objectives over all FPs is to support scientific work of the highest quality. In this context, the European scientific community should benefit from mobile researchers being involved in transnational projects (see CORDIS 2006).

---

[4] See Roediger-Schluga and Barber (2006) for a discussion on the history and different scopes of the Framework Programmes.

The study at hand draws on information on joint R&D projects funded in FP5[5]. For the interpretation of the results in the current study it is worth pointing to some important governance structures of FP5. Project proposals are to be submitted by self-organised consortia (see European Council 1998). FP5, with its corresponding financial support, is open to all legal entities established in the Member States of the European Union – individuals, industrial and commercial firms, universities, research organisations, etc. Proposals can be submitted by at least two independent legal entities established in different EU Member States or in an EU Member State and an associated State (see CORDIS 1998). Proposals to be funded are selected on the basis of criteria including scientific excellence, added value for the European Community, the potential contribution to furthering the economic and social objectives of the Community, the innovative nature, the prospects for disseminating/exploiting the results, and effective trans-national cooperation.

# 4 Observing R&D Collaborations across European Regions: Data and some Descriptive Statistics

This section discusses in some detail the data used and the empirical setting, including the construction of the dependent variables that will be implemented in the spatial interaction model. The first step in analysing the geography of industry and public research R&D networks is to construct region-by-region collaboration matrices. Our core data set to capture collaborative activities in Europe is the *sysres EUPRO* database[6], which presently comprises data on funded research projects of the EU FPs (complete for FP1-FP5, and about 70% complete for FP6) and all participating organisations. It contains systematic information on project objectives and achievements, project costs, project funding and contract type as well as on the participating organisations including the full name, the full address and the type of the organisation. We use a concordance scheme between postal codes and NUTS regions provided by Eurostat to trace the specific NUTS-2 region of an organisation. Thus, the *sysres EUPRO* database represents a valuable resource not only for this study, but for any kind of empirical analysis on the geography of knowledge creation and diffusion across Europe.

---

[5] FP5 had a total budget of 13.7 billion EUR and ran from 1998-2002 (CORDIS 1998). It focused on a limited number of research areas combining technological, industrial, economic, social and cultural aspects. The thematic priorites in FP5 are the following (Subprogramme name given in brackets): Quality of Life and management of living resources (Quality of Life); User-friendly information society (IST); Competitive and sustainable growth (GROWTH); Energy, environment and sustainable developement (EESD); Confirming the international role of community research (INCO2); Promotion of innovation and encouragement of SME participation (Innovation/SMEs); Improving the human reserach potential and the socio-economic knowledge base (Improving) (CORDIS 1998). It is worth noting that FP5 emphasised the protection of intellectual property rights in order to improve the efficiency of collaboration within the various types of European research projects. Project management procedures were reorganised. For instance, the role of project coordinators became increasingly important, because new administrative duties with respect to scientific coordination, supervision, and reports on financial issues were now assigned to them. Details on the formal application procedures are given by the European Council (1998).

For the construction of the region-by-region matrices, we use data on joint research projects funded within FP5. The European coverage in this study is achieved by using cross-section data on $i, j = 1, \ldots, n = 255$ NUTS-2 regions (NUTS revision 2003) of the 25 pre-2007 EU member-states, as well as Norway and Switzerland (see Appendix A for an explanation of the NUTS acronym, and the detailed list of regions). Although varying considerably in size, NUTS-2 regions are widely viewed as the most appropriate unit for modelling and analysis purposes since the EU cohesion funds are allocated at the level of NUTS-2, and have been used in similar empirical studies (see, for example, Fingleton 2001, Maurseth and Verspagen 2002, Fischer et al. 2006, LeSage et al. 2007).

To compare the geographical dimension of industry and public research networks, we have to construct two $n$-by-$n$ matrices labelled $P^{(\text{ind})}$ for intra-industry collaborations and $P^{(\text{edu})}$ for public research collaborations. The construction of $P^{(\text{ind})}$ and $P^{(\text{edu})}$ is based on a two-step procedure: *First*, for $P^{(\text{ind})}$ we extract all FP5 projects from the *sysres EUPRO* database that exclusively consist of industry participants (797 collaborative research projects with 1697 participants), while for $P^{(\text{edu})}$ we extract all

---

[6] The *sysres EUPRO* database is constructed and maintained by the *AIT Foresight & Policy Development Department* by substantially standardising raw data on FP research collaborations obtained from the CORDIS database (see Roediger-Schluga and Barber 2008).

FP5 projects from the *sysres EUPRO* database that exclusively consist of public research organisations (2184 collaborative research projects with 3423 participants). *Second*, we aggregate the number of individual collaborative activities to the regional level which leads to the observed number of R&D collaborations between two regions $i$ and $j$. We follow full counting procedure[7], i.e. for a project with three participating organisations in three different regions, say regions *a*, *b*, and *c*, we count three links: from region *a* to region *b*, from *b* to *c* and from *a* to *c*. When all three participants are located in one region we count three intraregional links. Note that we have excluded self loops to eliminate artificial self collaborations.

**Table 1: Some descriptive statistics on R&D collaborations among European regions as captured by joint EU FP5 research projects**

|  | Sum | Mean | SD | Min | Max | Skewness | Kurtosis |
|---|---|---|---|---|---|---|---|
| **Intra-industry cross-region collaborations $P^{(ind)}$** | 12,554 | 0.19 | 1.39 | 0 | 152 | 39.37 | 3,006.35 |
| **Intra-public cross-region collaborations $P^{(edu)}$** | 75,980 | 1.16 | 5.22 | 0 | 696 | 44.51 | 4,979.07 |

As a prelude to the analysis the follows, Table 1 presents some descriptive statistics on R&D collaborations among European regions as captured by joint EU FP5 research projects. It can be seen that the number of observed intra-public-research collaborations is about six times as high as the number of observed intra-industry collaborations. The mean interaction intensity between all regions for the public research sector is 1.16, while for the industry sector it is 0.19. However, the distribution of both industry and public research interactions show some similar characteristics. The standard deviation is much higher than the mean. The statistics for skewness and kurtosis point to an extremely right-skewed distribution, i.e. the main characteristic of the distribution of both *n*-by-*n* collaboration matrices is the fact that there are relatively few region pairs with a high number of collaborations, while the majority of the region pairs show relatively low collaboration intensities.

---

[7] Another counting method would be fractional counting by dividing each link in a project by the total number of links in a project. The full counting procedure used in the current study overestimates the impact of large projects, while the impact of large projects would be underestimated using the fractional counting method. We prefer the full counting method since full rather than fractional counting does justice to the true integer nature of R&D collaborations and is applicable in the context of a Poisson model specification.

# 5 Spatial Structure of Industry and Public Research Collaborations

This section sheds some light on the spatial distribution of European R&D networks in the industry sector and the public research sector. The spatial structure will be analysed by means of geographic maps, and through statistics from network analysis. As a starting point, Table 2 gives an overview of the Top-10 interregional industry and public research collaboration flows in FP5. It is quite striking that the region of Île-de-France is part of nearly all Top-10 region pairs in both sectors. The highest interaction intensity in the industry sector is observed for the regions of Île-de-France and Lombardia (31 collaborations). In the public research sector the region pair of Île-de-France and Denmark has the highest collaboration intensity (177 collaborations)[8].

**Table 2: Top-10 Interregional Collaboration Flows in the 5th EU FP**

| Intra-Public Research Collaborations $P^{(edu)}$ | | Intra-Industry Collaborations $P^{(ind)}$ | |
|---|---|---|---|
| **Region Pair** | **Number of Collaborations** | **Region Pair** | **Number of Collaborations** |
| Île-de-France and Denmark | 177 | Île-de-France and Lombardia | 31 |
| Île-de-France and Inner London | 130 | Oberbayern and Lombardia | 26 |
| Île-de-France and East Anglia | 128 | Île-de-France and Rhone Alpes | 25 |
| Île-de-France and Karlsruhe | 121 | Île-de-France and Cologne | 25 |
| Île-de-France and Oberbayern | 120 | Île-de-France and Outer London | 23 |
| Île-de-France and Darmstadt | 113 | Île-de-France and Oberbayern | 21 |
| Île-de-France and Piemonte | 107 | Outer London and Brussels | 21 |
| Île-de-France and Athens | 102 | Outer London and Denmark | 20 |
| Île-de-France and Noord-Brabant | 102 | Île-de-France and Düsseldorf | 20 |
| Denmark and Etelä-Suomi | 101 | Île-de-France and Darmstadt | 19 |

Figure 1 illustrates the spatial network of cross-region intra-industry and public research collaborations. Note that the region-by-region network is an undirected graph from a network analysis perspective. The nodes represent regions; their size is relative to their degree centrality corresponding to the number of links connected to a region. The spatial network maps reveal a quite different spatial structure of industry and public research networks across European regions. It is notable that collaborative activities in the industry sector are obviously more clustered in the centre of Europe than public research collaborations. Intra-public-research collaborations cover a much wider spatial area. Furthermore, the maps indicate that some regions playing no or only a small role in the industry sector, show high collaboration activities in the public research sector, such as for instance the regions of Lisbon, Madrid, Catalonia, Lazio, the capital regions of Greece and regions of the Eastern European countries. Île-de-France is the central hub in both spatial networks. A high density in both cases can also be observed for southeastern regions of the UK, northern Italian regions, southern and western regions in Germany, the Netherlands, Belgium and Switzerland.

---

[8] Note that the NUTS-2 level is equal to the NUTS-0 level, i.e. the country level, for the case of Denmark.

**Figure 1: Cross-region R&D collaborations in Europe.
A Industry Sector, B Public Research Sector**

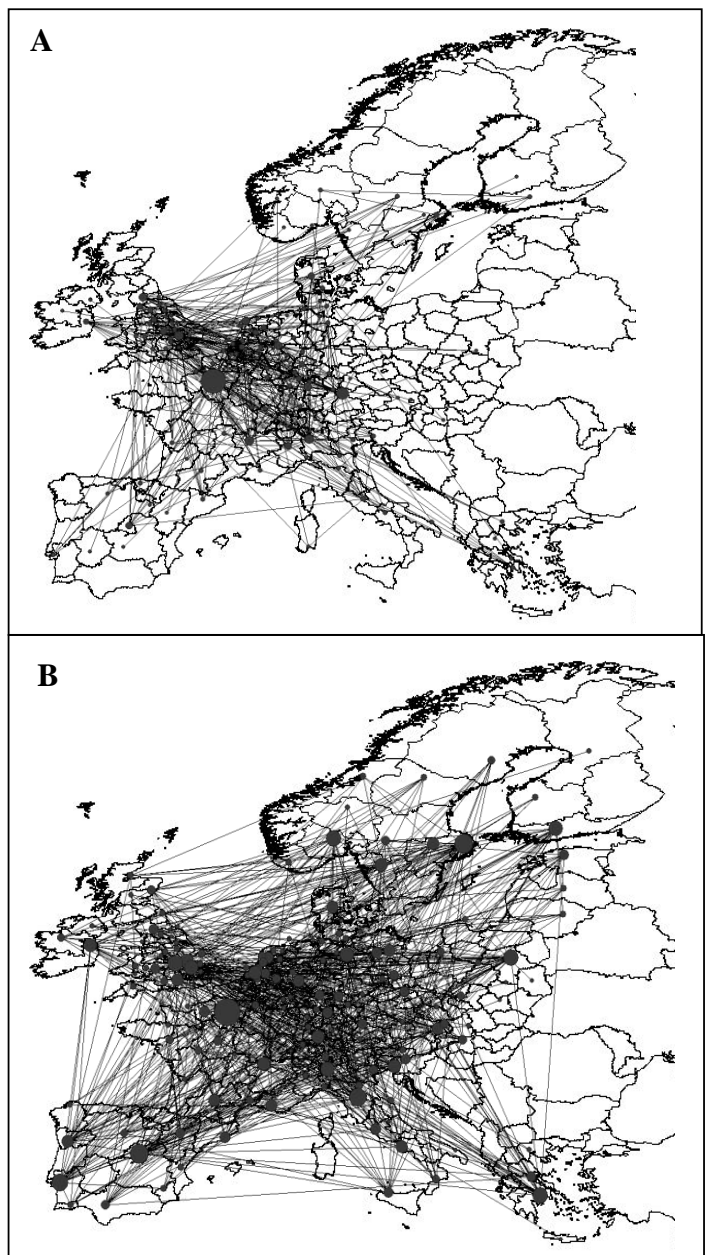

Note: Node size corresponds to a region's degree centrality
(see Wasserman and Faust 1994)

Table 2 and Figure 1 indicate the central regional players in the European FPs with respect to the number of project participations (i.e. degree centrality). We additionally calculate the eigenvector centrality, a more sophisticated version of the same idea. Having many connections surely affords influence and power, but not all connections are the same. Typically, connections to regions who are themselves well connected (high degree) will provide regions with more influence than connections to poorly connected (low degree) regions. Eigenvector centrality thus accords each region a centrality that depends both on the number and the quality of its connections by examining all regions in parallel and assigning centrality weights that correspond to the average centrality of all linked regions (see Bonacich 1987). The centrality values are scaled to the interval [0, 1], with higher values indicating greater centrality. This is done in each case using the region-by-region collaboration matrices $\boldsymbol{P}^{(ind)}$ and $\boldsymbol{P}^{(edu)}$, producing a vector of centralities that is an eigenvector of $\boldsymbol{P}^{(ind)}$ and $\boldsymbol{P}^{(edu)}$ corresponding to the largest eigenvalue. The eigenvector centrality is mathematically natural because it is based on fundamental properties of $\boldsymbol{P}^{(ind)}$ and $\boldsymbol{P}^{(edu)}$, specifically, their spectral properties.

Figure 2 visualises the eigenvector centrality of European regions in industry and public research R&D networks, while Table 3 gives the Top-10 regions by their eigenvector centrality in the industry and public research sector networks, accompanied by the number of participations per organisation and the contribution of large organisations to a region's connectivity. In both sectors the region of Île-de-France has by far the highest eigenvector centrality, i.e. Île-de-France has not only the highest number of connections but is also very well connected to other regions that have a high centrality. The gap in centrality between Île-de-France and the other regions is remarkable. Lombardia (rank 4 in the public sector, rank 2 in the industry sector) and Denmark (8 and 10, respectively) are the only regions that are also represented in both sectors. All other Top-10 regions of the public research sector are not placed in the Top-10 of the industry sector, and vice versa. It is noteworthy that capital regions of southern European countries show a very high eigenvector centrality in the public research sector (Athens, Madrid and Lazio), while they play a minor role in the industry sector. As visualised in Figure 2, this applies also for Eastern European regions, in particular the Baltic countries.

**Figure 2:** Eigenvector centrality of European regions in European R&D networks. A Industry Sector, B Public Research Sector

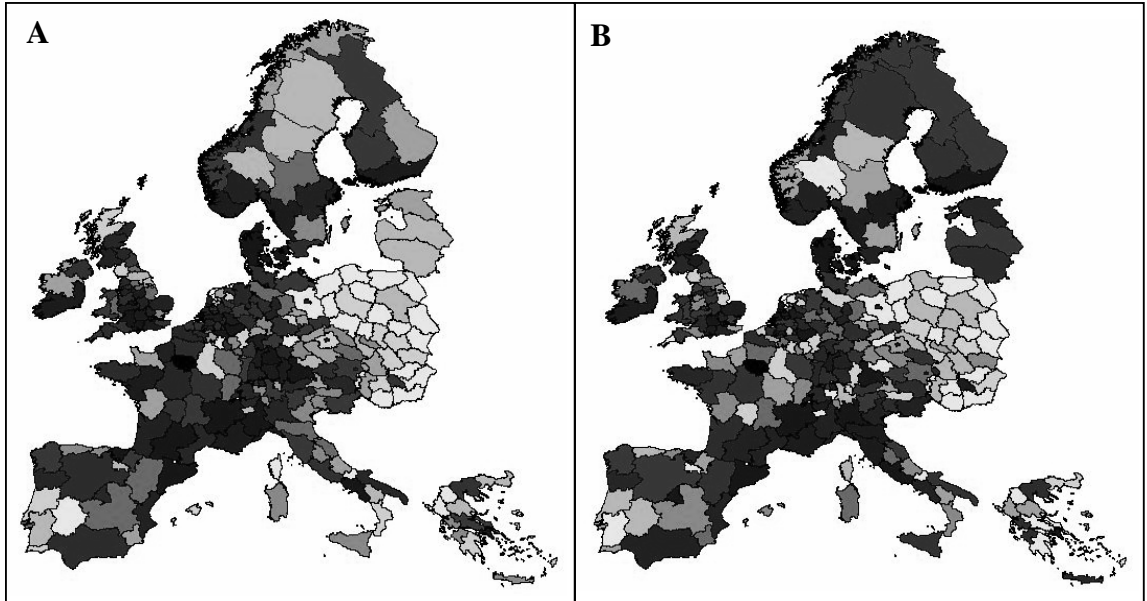

Note: Regional eigenvector centrality is visualised by means of grey scales (the darker the shading, the higher the region`s eigenvector centrality).

An interesting question in this context is whether the high connectivity of the regions listed in Table 3 depends on a high number of regional participants, i.e. does the high eigenvector centrality of these regions depend on the role of a few strong players or on diffused participation activities by a large number of participating organisations? The figures on organisational participation intensity and the contribution of large firms to total FP participation of Table 3 shed some light on these issues. It can be seen that the role of strong players varies considerably across regions. In the industry sector, large players play a very important role in the regions Oberbayern, Stuttgart, Piemonte and Île-de-France by contributing more or nearly 50% to total regional FP5 participation (see third column of Table 3)[9]. This is also reflected by the organisational participation intensity in these regions. In contrary, the region of Karlsruhe is characterised by a very diffused participation activity since large organisations contribute only about 3.41% to the connectivity of this region, also reflected by a lower organisational participation intensity. In the public research sector, the regions Île-de-France and Noord-Holland are characterized by a high contribution of large players in public research FP networks[10], while in Danmark and Caluna participation activity is more dispersed across different organisations.

---

[9] These large players refer to the organisations Siemens AG and BMW AG located in the region Oberbayern, Daimler Chrysler AG and Robert Bosch AG located in the region Stuttgart, Fiat Gruppo and Telecom Italia located in the region Piemonte, and Thales Group, EADS and Alcatel-Lucent located in the region Île-de-France.

**Table 3: Top-10 regions by means of Eigenvector centrality (FP5)**

| Region | Eigenvector Centrality | Organisational participation intensity* | Participation intensity from large organisations (%)** |
|---|---|---|---|
| **Intra-Industry Collaborations $P^{(ind)}$** | | | |
| Île-de-France | 1.000 | 3.18 | 46.63 |
| Lombardia | 0.278 | 1.72 | 17.82 |
| Oberbayern | 0.212 | 3.06 | 54.23 |
| Piemonte | 0.205 | 3.24 | 55.81 |
| Outer London | 0.200 | 1.97 | 22.77 |
| Stuttgart | 0.189 | 3.54 | 51.14 |
| Noord-Brabant | 0.173 | 2.24 | 34.44 |
| Midi-Pyrénées | 0.169 | 2.75 | 32.28 |
| Karlsruhe | 0.154 | 1.51 | 3.41 |
| Denmark | 0.127 | 1.77 | 15.22 |
| **Intra-Public Research Collaborations $P^{(edu)}$** | | | |
| Île-de-France | 1.000 | 5.91 | 39.66 |
| Lazio | 0.523 | 4.81 | 22.35 |
| Athens | 0.415 | 7.62 | 25.70 |
| Lombardia | 0.432 | 5.51 | 18.62 |
| Madrid | 0.411 | 5.56 | 20.74 |
| Cataluna | 0.401 | 4.14 | 6.74 |
| Noord-Holland | 0.398 | 10.01 | 37.42 |
| Denmark | 0.391 | 5.42 | 9.04 |
| Inner London | 0.388 | 6.43 | 32.63 |
| Etelä-Suomi | 0.385 | 6.93 | 20.53 |

Note: Eigenvector centrality is normalized between zero and one.
* Organisational participation intensity is defined as the number of participations per organisation ** By large organisations we mean organisations in the 99th percentile with respect to the number of project participations in FP5.

# 6  The spatial interaction approach

We will measure separation effects on the constitution of cross-region intra-industry and cross-region intra-public-research R&D collaborations in Europe. We rely on the spatial interaction model of the type used by Scherngell and Barber (2009) in a similar context. Spatial interaction models incorporate a function characterising the origin *i* of interaction, a function characterising the destination *j* of interaction and a function characterising the separation between two regions *i* and *j*. The model is characterised by a formal distinction implicit in the definitions of origin and destination functions on the one hand, and separation functions on the other (see, for example, Sen and Smith 1995). Origin and destination functions are described using weighted origin and destination variables, respectively, while the separation functions are postulated to be explicit functions of numerical separation variables. The general model is given by

$$P_{ij} = A_i \ B_j \ S_{ij} + \varepsilon_{ij} \qquad\qquad i, j = 1, \ldots, n \qquad (1)$$

where $P_{ij}$ reflects the number of observed collaboration flows between region *i* and region *j*. The spatial interaction approach described in this section is equally applicable to $P^{(ind)}$ and $P^{(edu)}$. We do not distinguish between them in the formal presentation. $A_i$ denotes the origin function, $B_j$ denotes the destination function, while $S_{ij}$ represents a separation function. Based on the general model we distinguish three model versions: Model (a) uses, for purposes of comparison, the total cross-region FP5 R&D collaborations as dependent variable as in Scherngell and Barber (2009); model (b) uses the intra-industry cross-region FP5 R&D collaborations as given by $P^{(ind)}$; and model (c) uses the intra-public-research cross-region FP5 R&D collaborations as given by $P^{(edu)}$. We define $A_i = A(a_i, \alpha_1) = a_i^{\alpha_1}$, and $B_j = B(b_j, \alpha_2) = b_j^{\alpha_2}$. In Model (a), $a_i$ and $b_j$ are simply measured in terms of the number of organisations participating in EU FP5 projects in the region *i* and *j*, respectively, while for Model (b) we only use the number of firms, and in Model (c) the number of public research organisations. $\alpha_1$ and $\alpha_2$ are scalar parameters to be estimated. Based on the model specification the product of the functions $A_i B_j$ can be simply interpreted as the number of cross-region R&D collaborations which are possible. Note that due to symmetry of the origin and destination variables we have a special case with [$\alpha_1 = \alpha_2$], i.e. numerical results for $\alpha_1$ and $\alpha_2$ should be equal up to numerical precision.

---

[10] These large players refer to the organisations Centre National de la Recherche Scientifique (CNRS) located in the region Île-de-France, and the Unversiteit Amsterdam and the Vrije Universiteit Amsterdam located in the region Noord-Holland.

The separation function $S_{ij}$ is of central importance by nature of our research questions. We use a multivariate exponential functional as given by

$$S_{ij} = \exp\left[\sum_{k=1}^{K} \beta_k \, d_{ij}^{(k)}\right] \qquad i,j = 1,\ldots,n \qquad (2)$$

where $d_{ij}^{(k)}$ are $K$ separation measures and $\beta_k$ ($k = 1, \ldots, K$) are parameters to be estimated that will show the relative strengths of the separation measures. We focus on $K = 7$ distinct measures of separation. In accord with our main lines of hypotheses, we can group these variables into three categories:

(i) Variables accounting for *spatial effects*: $d_{ij}^{(1)}$ denotes geographical distance between two regions $i$ and $j$ as measured by the great circle distance between the economic centres of the regions. $d_{ij}^{(2)}$ and $d_{ij}^{(3)}$ are dummy variables that control for neighbouring region- and neighbouring country effects, respectively.

(ii) Variables accounting for *institutional and cultural effects*: $d_{ij}^{(4)}$ is a country border dummy variable that takes a value of zero if two regions $i$ and $j$ are located in the same country, and one otherwise, while $d_{ij}^{(5)}$ is a language area dummy variable that takes a value of zero if two regions $i$ and $j$ are located in the same language area, and one otherwise.

(iii) Variables accounting for *sectoral effects*: $d_{ij}^{(6)}$ measures technological distance by using regional patent data from the European Patent office (EPO). The variable is constructed as a vector $t(i)$ that measures region $i$'s share of patenting in each of the technological subclasses of the International Patent Classification (IPC). Technological subclasses correspond to the third-digit level of the IPC systems. We use the Pearson correlation coefficient given by $r^2 = corr[t(i), t(j)]^2$ between the technological vectors of two regions $i$ and $j$ to define how close they are to each other in technological space. Their technological distance is given by $d_{ij}^{(6)} = 1 - r^2$. $d_{ij}^{(7)}$ measures the economic similarity between regions. We use data on regional Gross Value Added (GVA) in five different sectors (agriculture, manufacturing, construction, market services and non-market services) from the Cambridge

Econometrics database to construct a vector *g*(*i*) that measures region *i*'s share of GVA in the five sectors. We again rely on the Pearson correlation coefficient and define $d_{ij}^{(7)} = corr[g(i), g(j)]$.

Incorporating the origin, destination and separation functions into Equation (1) leads to the empirical model:

$$P_{ij} = a_i^{\alpha_1} b_j^{\alpha_2} \exp\left[\sum_{k=1}^{K} \beta_k \ d_{ij}^{(k)}\right] + \varepsilon_{ij}. \tag{3}$$

We are interested in estimating the parameters $\alpha_1 = \alpha_2$ and $\beta_k$, for model (a) using all FP5 collaborations, model (b) using intra-industry FP5 collaborations, and model (c) using intra-public-research collaborations as dependent variable. OLS estimation procedures are not appropriate since we model count data. Thus, it is natural to use a Poisson model specification given by:

$$f(p_{ij}) = V_{ij}^{p_{ij}} \ e^{-V_{ij}} \Big/ p_{ij}! \tag{4}$$

where $V_{ij} = A_i \ B_j \ S_{ij}$. However, estimation results in Scherngell and Barber (2009) show that unobserved heterogeneity which cannot be captured by the covariates leads to biased estimates due to overdispersion (i.e. the variance is higher than the expectation). This problem is solved by introducing a stochastic heterogeneity parameter exp($\xi_{ij}$), leading to a Negative Binomial density distribution that is given by (see Long and Freese 2001)

$$f(P_{ij}) = \frac{\Gamma(P_{ij} + \delta^{-1})}{\Gamma(P_{ij}+1)\Gamma(\delta^{-1})} \left(\frac{\delta^{-1}}{A_i \ B_j \ S_{ij} + \delta^{-1}}\right)^{\delta^{-1}} \left(\frac{A_i \ B_j \ S_{ij}}{A_i \ B_j \ S_{ij} + \delta^{-1}}\right)^{p_{ij}} \tag{5}$$

where $\Gamma(\cdot)$ denotes the gamma function. Model estimation is done by Maximum Likelihood procedures (see Cameron and Trivedi 1998).

# 7 Estimation results

Table 4 presents the sample estimates of the spatial interaction models, with standard errors given in brackets. We use the Negative Binomial model specification as given by Equation (5). The dispersion parameter $\delta$ is significant for all three model versions, indicating that the Negative Binomial specification is the right specification, i.e. the standard Poisson specification would be biased due to unobserved heterogeneity between the region pairs. The existence of unobserved heterogeneity that cannot be captured by the covariates leads to overdispersion and, thus, to biased model parameters for the standard Poisson model.

In the context of the relevant literature on innovation and knowledge diffusion, and the hypotheses proposed in Section 2, our model produces some interesting results. The estimation results of model (a) provide evidence that geographical distance between two organisations has a significant negative effect on the likelihood that they collaborate. The parameter estimate of $\beta_1 = -0.278$ indicates that for each additional 100 km between two organisations, the mean collaboration frequency decreases by 25.6%. The geographic localisation of collaboration activities in the FPs is also reflected by the estimates for $\beta_2$ and $\beta_3$. The likelihood of collaboration between two organisations increases when they are located in neighbouring regions or neighbouring countries, respectively. However, these effects seem to be quite heterogeneous with respect to the two sectors that are in the focus this study. In the industry sector spatial separation effects are much more important then in the public research sector as evidenced by the results of model (b) and model (c). In Model (b) the estimate for $\beta_1$ increases in magnitude by 68% to a value of -0.404. In contrast, in model (c) $\beta_1$ is estimated with a value of -0.069, i.e. geographical distance has a much weaker effect for intra-public-research collaborations than for intra-industry networks. The effect of geographical distance nearly disappears in the case of intra-public-research collaboration activities. Also, neighbouring region and neighbouring country effects are lower for public research networks. In the context of the objectives of the FPs regarding the integration of the European research area, these results indicate that there remains still much space for improvement in the industry sector, while in the public research sector barriers of geography now provide only a small hurdle.

**Table 4: Estimation Results of the Negative Binomial Spatial Interaction Models**
[65,025 observations, asymptotic standard errors given in brackets]

|  | Negative Binomial spatial interaction model | | |
|---|---|---|---|
|  | **(a) Total FP5** | **(b) FP5 Industry** | **(c) FP5 Public Research** |
| ***Origin and destination variable*** [$\alpha_1 = \alpha_2$] | 0.706*** (0.003) | 0.878*** (0.020) | 0.897*** (0.013) |
| ***Spatial effects*** | | | |
| *Geographical distance* [$\beta_1$] | -0.278*** (0.008) | -0.404*** (0.034) | -0.069*** (0.011) |
| *Neighbouring region* [$\beta_2$] | 0.184*** (0.036) | 0.254** (0.071) | 0.132*** (0.084) |
| *Neighbouring country* [$\beta_3$] | 0.281*** (0.014) | 0.111* (0.051) | 0.103* (0.041) |
| ***Institutional and cultural effects*** | | | |
| *Country border* [$\beta_4$] | -0.008 (0.023) | -0.023 (0.052) | 0.017 (0.015) |
| *Language area* [$\beta_5$] | -0.098*** (0.024) | -0.271*** (0.091) | -0.023* (0.007) |
| ***Sectoral effects*** | | | |
| *Technological distance* [$\beta_6$] | -1.413*** (0.115) | -1.532*** (0.213) | -0.761*** (0.105) |
| *Economic similarity* [$\beta_7$] | 0.534*** (0.042) | 1.300*** (0.278) | -0.092 (0.159) |
| **Constant** | -2.539*** (0.128) | -11.694*** (0.562) | -7.683*** (0.632) |
| **Dispersion parameter** ($\delta$) | 1.047*** (0.009) | 0.544*** (0.037) | 0.470*** (0.015) |
| **Log-Likelihood** | -135,234.21 | -112,257.87 | -125,741.23 |
| **Sigma Square** | 6.523 | 7.003 | 6.923 |

Notes: The dependent variable in model (a) is the cross-region collaboration intensity between two regions *i* and *j*, in model (b) the cross-region intra-industry collaboration intensity between two regions *i* and *j*, and in model (c) the cross-region intra-public-research collaboration intensity between two regions *i* and *j*. The independent variables are defined as given in the text. Note that we tested the residual vector for the existence of spatial autocorrelation which could be a problem in the context of interaction data (see LeSage, Fischer and Scherngell 2007). The respective Moran´s *I* statistic is insignificant, i.e. spatial autocorrelation in the error term does not exist. ***significant at the 0.001 significance level, **significant at the 0.01 significance level, *significant at the 0.05 significance level

Country borders – as evidenced by the estimate for $\beta_4$ – are not significant in any model version. This indicates that one of the main EU objectives, namely to overcome the institutional barriers and encourage cross-country collaboration, has been met[11]. This is consistent with findings of Almendral et al. (2007). However, the estimate for $\beta_5$ tells us that in the industry sector R&D collaborations are hampered by cultural barriers – firms that are located in the same language area show a higher probability to collaborate – and the EU FPs have not yet eliminated such effects in the industry sector. In contrast, in the public research sector language barriers seem to play only a very small role, at least concerning FP-funded joint research collaborations.

For all model versions, most important are technological distance effects as evidenced by the parameter estimates for $\beta_6$. This implies that it is most likely that cross-region R&D collaborations occur between regions that are close to each other in technological space. This finding is in line with previous results of Fischer et al. (2006) and LeSage et al. (2007) for the case of interregional knowledge spillovers, but the technological distance effect they found is much higher than in the current study for interregional FP collaborations. However, it is notable that technological distance effects are much higher in the industry sector than in public research. Economic similarity between regions is very important for the industry sector, while insignificant for the public research sector as indicated by the parameter estimates for $\beta_7$. As expected, the estimates for the origin and destination variables indicate that a higher number of participating organisations in a region increases the probability of collaboration with other regions.

## 8 Concluding remarks

One of the key current research fields in economic geography and economics of innovation is the empirical analysis of the geography of innovation. In particular, the geographical dimension of phenomena such as R&D collaborations is of special interest in order to gain insight into the spatial diffusion of knowledge. The analysis of the geography of R&D collaborations has important policy implications for the EU, for instance with respect to the spatial scale of innovation systems and R&D interactions.

---

[11] Most likely this is largely due to the governance rule that each project must have international partners.

The focus of this study has been on investigating the role of geographical space for R&D collaborations funded by the 5$^{th}$ EU FP from a regional perspective. The objective was to identify separation effects on the constitution of cross-region R&D collaborations. We used an appropriate analytical framework, the Negative Binomial spatial interaction model, to estimate the separation effects.

The study produces some promising results in the context of the empirical literature on innovation. Geographical distance and co-localisation of organisations in neighbouring regions are important determinants of the constitution of cross-region R&D collaborations in Europe. However, model estimations show that these geographical effects are much higher for intra-industry cooperative activities than for collaborations between public research organisations, where negative effects of geography nearly vanish. For all model versions, technological proximity is more important than spatial effects. R&D collaborations occur most often between organisations that are not too far from each other in technological space. R&D collaborations are also determined by language barriers, but language barrier effects are smaller than geographical effects. Country border effects are not observable.

In a European technology policy context, the results provide significant implications. Concerning one of the main policy objectives, namely the integration of research efforts at the level of the EU, the results of the study at hand point to mixed policy outcomes: *First*, The European FPs seem to help overcoming geographical distance barriers in the public research sector, while in the industry sector these barriers still play an important role. *Second*, one important achievement of the FPs seems to be reduction of institutional barriers – as captured by country border effects – for research collaboration between organisations in Europe, both in the industry and the public research sector. *Third*, concerning language area effects – used as a proxy for cultural barriers – there is much space for improvement, in particular in the industry sector. *Fourth*, it seems that the objective to foster inter-technology research collaborations in the FPs remains far from realisation, as indicated by the influence of technological distance.

The study raises some topics for future research. *First*, the estimation of this model for different FPs would shed some light on the temporal evolution of these effects and provide some insight into the integration of R&D collaborations with respect to geography and technology over time. A *second* promising research direction would be to apply the spatial interaction model for cross-region collaborations in different subprogrammes. This may provide insight on the role of the separation variables across different scientific fields. *Third*, the definition of the region-by-region collaboration matrix is based on the assumption that all project participations are equally important. This is of course an approximation to reality. Thus, other counting methods could be explored. *Fourth*, the present work would be well complemented by an investigation of science-industry interactions utilising data on joint research projects between firms and public research organisations.

**Acknowledgements.** The work presented here is partly funded by the EU-FP6-NEST project NEMO ('Network Models, Governance and R&D collaboration networks'), contract number 028875. We thank two anonymous referees. Their comments significantly improved an earlier version of the current study.

**Appendix A**

NUTS is an acronym of the French for the "nomenclature of territorial units for statistics", which is a hierarchical system of regions used by the statistical office of the European Community for the production of regional statistics. At the top of the hierarchy are NUTS-0 regions (countries) below which are NUTS-1 regions and then NUTS-2 regions. This study disaggregates Europe's territory into 255 NUTS-2 regions located in the EU-25 member states (except Cyprus and Malta) plus Norway and Switzerland. We exclude the Spanish North African territories of Ceuta y Melilla, the Portuguese non-continental territories Azores and Madeira, and the French Departments d'Outre-Mer Guadeloupe, Martinique, French Guayana and Reunion. Thus, we include the following NUTS 2 regions:

| | |
|---|---|
| *Austria*: | Burgenland; Niederösterreich; Wien; Kärnten; Steiermark; Oberösterreich; Salzburg; Tirol; Vorarlberg |
| *Belgium*: | Région de Bruxelles-Capitale/Brussels Hoofdstedelijk Gewest; Prov. Antwerpen; Prov. Limburg (BE); Prov. Oost-Vlaanderen; Prov. Vlaams-Brabant; Prov. West-Vlaanderen; Prov. Brabant Wallon; Prov. Hainaut; Prov. Liége; Prov. Luxembourg (BE); Prov. Namur |
| *Czech Republic*: | Praha, Stredni Cechy, Jihozapad, Severozapad, Severovychod, Jihovychod, Stredni Morava, Moravskoslezsko |
| *Denmark*: | Danmark |
| *Estland*: | *Eesti* |
| *Finland*: | Itä-Suomi; Etelä-Suomi; Länsi-Suomi; Pohjois-Suomi |
| *France*: | Île de France; Champagne-Ardenne; Picardie Haute-Normandie; Centre; Basse-Normandie; Bourgogne; Nord-Pas-de-Calais; Lorraine; Alsace; Franche-Comté; Pays de la Loire; Bretagne; Poitou-Charentes; Aquitaine; Midi-Pyrénées; Limousin; Rhône-Alpes; Auvergne; Languedoc-Roussillon; Provence-Côte d'Azur; Corse |
| *Germany*: | Stuttgart; Karlsruhe; Freiburg; Tübingen; Oberbayern; Niederbayern; Oberpfalz; Oberfranken; Mittelfranken; Unterfranken; Schwaben; Berlin; Brandenburg; Bremen; Hamburg; Darmstadt; Gießen; Kassel; Mecklenburg-Vorpommern; Braunschweig; Hannover; Lüneburg; Weser-Ems; Düsseldorf; |
|  | Köln; Münster; Detmold; Arnsberg; Koblenz; Trier; Rheinhessen-Pfalz; Saarland; Chemnitz; Dresden; Leipzig; Dessau; Halle; Magdeburg; Schleswig-Holstein; Thüringen |
| *Greece*: | Anatoliki Makedonia; Kentriki Makedonia; Dytiki Makedonia; Thessalia; Ipeiros; Ionia Nisia; Dytiki Ellada; Sterea Ellada; Peloponnisos; Attiki; Voreio Aigaio; Notio Aigaio; Kriti |
| *Hungary*: | Kuzup-Magyarorszßg, Kuzup-Dunssnt, Nyugat-Dunssnt, Dus-Dunsst, Oszak-Magyarorszßg, Oszak-Alfald, Dus-Alfad |
| *Ireland*: | Border, Midland and Western; Southern and Eastern |
| *Italy*: | Piemonte; Valle d'Aosta; Liguria; Lombardia; Trentino-Alto Adige; Veneto; Friuli-Venezia Giulia; Emilia-Romagna; Toscana; Umbria; Marche; Lazio; Abruzzo; Molise; Campania; Puglia; Basilicata; Calabria; Sicilia; Sardegna |
| *Latvia*: | Latvia |
| *Lithuania*: | Liuteva |
| *Luxembourg*: | Luxembourg (Grand-Duché) |
| *Netherlands*: | Groningen; Friesland; Drenthe; Overijssel; Gelderland; Flevoland; Utrecht; Noord-Holland; Zuid-Holland; Zeeland; Noord-Brabant; Limburg (NL) |
| *Norway*: | Oslo og Akershus, Hedmark og Oppland, Sor-Ïstlandet, Agder og Rogaland, Vestlandet, Trondelag, Nord-Norge |
| *Poland*: | Lodzkie, Mazowieckie, Malopolskie, Slaskie, Lubelskie, Podkarpackie, Swietokrzyskie, Podlaskie, Wielkopolskie, Zachodniopomorskie, Lubuskie, Dolnoslaskie, Opolskie, Kujawsko-Pomorskie, Warminsko-Mazurskie, Pomorskie |
| *Portugal*: | Norte; Centro (P); Lisboa e Vale do Tejo; Alentejo |
| *Slovakia*: | Bratislavsky kraj, Zaspadny Slovensko, Stredny Slovensko, Vachodny Slovensko |
| *Slovenija*: | Slovenija |
| *Spain*: | Galicia; Asturias; Cantabria; Pais Vasco; Comunidad Foral de Navar; La Rioja; Aragón; Comunidad de Madrid; Castilla y León; Castilla-la Mancha; Extremadura; Cataluña; Comunidad Valenciana; Islas Baleares; Andalucia; Región de Murcia |
| *Sweden*: | Stockholm; Östra Mellansverige; Sydsverige; Norra Mellansverige; Mellersta Norrland; Övre Norrland; Småland med Öarna; Västsverige |
| *Switzerland:* | Region Ümanique, Espace Mittelland, Nordwestschweiz, Zürich, Ostschweiz, Zentralschweiz, Ticino |

*United Kingdom*: Tees Valley & Durham; Northumberland & Wear; Cumbria; Cheshire; Greater Manchester; Lancashire; Merseyside; East Riding & .Lincolnshire; North Yorkshire; South Yorkshire; West Yorkshire; Derbyshire & Nottingham; Leicestershire; Lincolnshire; Herefordshire; Shropshire & Staffordshire; West Midlands; East Anglia; Bedfordshire & Hertfordshire; Essex; Inner London; Outer London; Berkshire; Surrey; Hampshire & Isle of Wight; Kent; Gloucestershire; Dorset & Somerset; Conwall & Isles of Scilly; Devon; West Wales; East Wales; North Eastern Scotland; Eastern Scotland; South Western Scotland; Highlands and Islands; Northern Ireland